# Analyzing the Habitable Zones of Circumbinary Planets Using Machine Learning

Zhihui Kong[1], Jonathan H. Jiang[2], Remo Burn[3], Kristen A. Fahy[2], Zonghong Zhu[1]

1. Department of Astronomy, Beijing Normal University, Beijing, China
2. Jet Propulsion Laboratory, California Institute of Technology, Pasadena, USA
3. Max Planck Institute for Astronomy, Königstuhl 17, 69117 Heidelberg, Germany

Correspondence: Jonathan.H.Jiang@jpl.nasa.gov

## Abstract

Exoplanet detection in the past decade by efforts including NASA's Kepler and TESS missions has discovered many worlds that differ substantially from planets in our own Solar System, including more than 150 exoplanets orbiting binary or multi-star systems. This not only broadens our understanding of the diversity of exoplanets, but also promotes our study of exoplanets in the complex binary systems and provides motivation to explore their habitability. In this study, we investigate the Habitable Zones of circumbinary planets based on planetary trajectory and dynamically informed habitable zones. Our results indicate that the mass ratio and orbital eccentricity of binary stars are important factors affecting the orbital stability and habitability of planetary systems. Moreover, planetary trajectory and dynamically informed habitable zones divide planetary habitability into three categories: habitable, part-habitable and uninhabitable. Therefore, we train a machine learning model to quickly and efficiently classify these planetary systems.

## 1 Introduction

We have discovered more than four thousand exoplanets to date, many of which have structures that vary greatly from our own Solar System. Meanwhile, previous research suggested that sun-like stars are common in binary systems, where almost half of the sun-like stars have a companion (Raghavan et al. 2010; Moe & Di Stefano 2017) and the rate of occurrence of planets with $R_p \geq 6R_\oplus$ orbiting with $P_p \leq 300d$ host circumbinary are as common as those orbiting single stars (Armstrong et al. 2014). Therefore, study on the planetary orbits in binary systems can help us to expand the concept of exoplanetary systems and to understand the complex formation of planetary systems.

Based on the Open Exoplanet Catalogue, there are 231 exoplanets that have been found, of which 192 are circumstellar planets and 39 circumbinary planets. There are five Kepler circumbinary planets (CBPs), Kepler-16b, Kepler-47c, Kepler-452b, Kepler-1747b, and Kepler-1661b reside in the Habitable Zones (HZ) of their binary host stars (Laurance R et al. 2011, Orosz et al. 2019, Welsh et al. 2015, Kostov et al. 2016, Socia et al. 2020). These discoveries open up a realm of new opportunity for studying habitable planets in binary systems. Consequently, we will focus on analyzing the Habitable Zone (HZ) of CBPs in this paper.

We must begin by defining the HZ in binary systems, which does present some challenges. To study the dynamics and stability of the CBPs we must use N-body numerical simulation or analytic formula developed from Rabl & Dvorak (1988) by Holman & Wiegert (1999) since there is no analytic solution to the three-body system. In principle, the N-body simulation method uses regressions of three-or-more body systems that check the resulting dynamical stability of each trial set of parameters, however, this requires high computation cost. The second-order polynomial method, as noted in Holman & Wiegert's work, cannot accurately evaluate the stability of circumbinary planets, especially if a circumbinary planet moves into 'Islands of Instability', which is the unstable region of orbital resonance.

Defining habitable zones in binary systems is complicated even further, as we must account for the planet receiving two sources of radiation, possibly from different spectral types. Additionally, the gravitational interaction between the planet and the two stars may change as the distance from the planet to each star changes over time. Therefore, the amount and spectral composition of light reaching a potentially habitable planet can vary on relatively short timescales (Haghighipour & Kaltenegger 2013; Cuntz 2014).

The is concept of "dynamically informed habitable zones" has made it possible to study the habitability prospects of a planet orbiting a star in a binary system (Eggl et al. 2012). Then it was found that dynamically informed habitable zones extended into circumbinary orbit (Eggl et al. 2020). However, the evolution of life on a planet requires not only that the initial conditions of the planet are favorable to the evolution of life, but also that the planet subsequently remains habitable without interruption. In this paper, we consider the influence of the HZ on both binary stars and planetary orbits. For instance, because of the gravitational perturbations of the circumbinary on the planet, the orbit of the planet is stabilized in a circular region for a long time. (Lam & Kipping 2018). Moreover, we designed machine learning algorithms, Deep Neural Networks (DNNs), that can be used to quickly and accurately distinguish different HZ's orbital types.

In Section 2, we describe the circumbinary planet model and the dynamically informed habitable zones. In Section 3, we describe the network architecture of DNNs. In Section 4, We show the statistical distributions for different types of habitable zones and the performance of our ML algorithm. Finally, we summarize the significance of our work and future areas for development in Section 5.

## 2 Model and Method

In this chapter, we introduce two models, one is how to model CBPs via the N-body simulation package, REBOUND, which was first developed by Rein & Liu (2012). The other is how to define and calculate the HZ of binary systems.

**2.1 The model of circumbinary planet systems**

We use an IAS15 integrator which is a non-symplectic integrator with adaptive time-stepping (Rein & Spiegel 2015) to study the P-type orbit stability. In the REBOUND environment, we set up a circumbinary planetary system. The central binary consists of mass $m_A$ and $m_B$ with total mass $m_A + m_B = 1\ m_\odot$ and the mass fraction of the binary $\mu$, where

$$\mu \equiv m_B/(m_A+m_B) \quad (1)$$

The binary orbit has initial distance, $d_b$, and the orbital eccentricity is $e_b$. The Keplerian orbit of the planet with mass $m_p$ around the center of mass of the binary, the initial eccentricity of the planet is zero, and the other orbital parameters are the semimajor axis, $a_p$, and inclination, $i$.

We ran $6*10^5$ REBOUND experiments with mass of the planet ($m_p = 0, m_\oplus, m_{jupiter}$), with each simulation lasting $10^3$ planetary period. For each simulation, we uniformly sampled the mass ratio, $\mu \in [0.2, 0.5]$, the initial binary eccentricity, $e_b[0,\ 0.8]$, and the inclination of the planet, $i \in [0, \pi]$. The initial distance of binary is set to 0.1 AU, and the initial eccentricity of the planet $e_p$ is set to zero. Note that our work focuses on the HZ of orbit-stabilized systems, therefore we threw away the extreme combination of parameters, likes $e_b > 0.8$ and $\mu < 0.2$. Table 1 illustrates the details.

Two conditions make circumbinary planetary systems unstable with the planet either colliding with the secondary star or being ejected from the system. The former instance can be tested by calculating the distance between the secondary star and the planet, if it is less than Hill's radius then a collision occurs. The latter can be tested by calculating the velocity, when the velocity exceeds the escape velocity, $v_{esc}$, the planet is ejected from the system. We perform these two test every quarter with a binary period and label the simulations as "instable" and subsequently terminate them. Then, we get two kinds of samples: "stable" and "instable". Next, we focus on the HZ of these stable samples. Figure 1 shows the planetary trajectory of a stable system

**Table 1:** Initial conditions for circumbinary planetary system

| Binary Parameters ||
|---|---|
| $m_A + m_B = 1\, m_\odot;\quad m_A > m_B;\quad d_b = 0.1 AU$ ||
| $0.2 \leq \mu \leq 0.5;\quad 0 \leq e_b \leq 0.8\ step\ width: 0.01$ ||
| Planet Parameters ||
| $0.5 d_b \leq a_p \leq 1.5 d_b\ step\ width: 0.01$ ||
| $m_p = 0,\quad m_\oplus,\quad m_{jupiter}$ ||
| Inclination: $i \in [0, \pi]$ | Step width: $\frac{\pi}{18}$ |
| Sample size: $6 * 10^5$ | $T = 10^3 T_P$ |

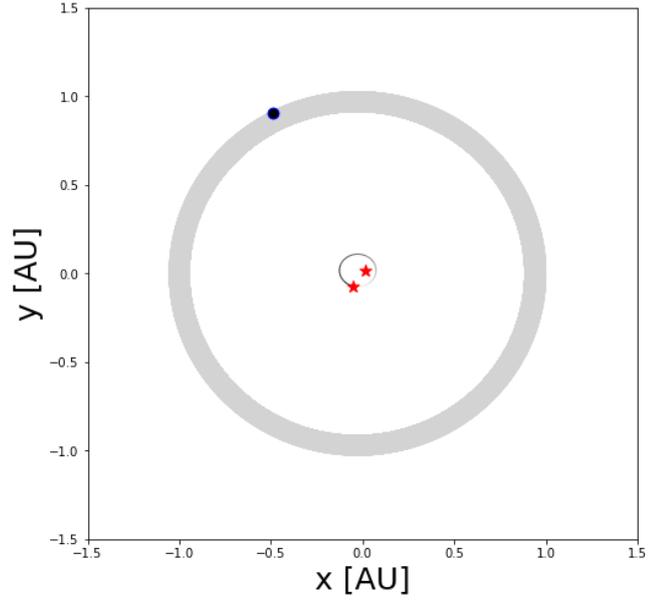

**Figure 1:** Shows the system after $10^3$ planetary periods. The red asterisk shows the binary stars and the blue dot is the planet; the light-gray region is the orbit trail of the planet. In this system, the mass fraction of the binary is $\mu = 0.2$; the orbital eccentricity is $e_b = 0.4$; the planet's mass is $m_p = 1 m_\oplus$ and initial semimajor axis is $a_p = 1 AU$.

## 2.2 Habitable zones of binary systems

Most research on habitability focuses on planetary systems that orbit a single star, in this case, the planet receives an almost constant amount of radiation on a permanent basis. As more and more planets in the binary systems are discovered, the study of HZ in the binary systems must interpret a broader view. A secondary star provides an additional source of radiation and gravitational perturbations. As a consequence, the radiation received by the planet in binary systems changes over time. How planetary atmospheres respond to this change is important in our work. To quantify the effects of planetary atmospheres in our work, we introduce "dynamically informed habitable zones". Firstly, we use the concept of climate inertia to simplify the model. Climate inertia refers to the time required for climate parameters, such as average surface temperature, to respond to radiative forcing. The faster the average surface temperature changes, the lower the climate inertia of the planet. The changes in insolation will quickly affect the planet's surface temperature when the climate inertia of the planet is small, therefore, the planet must remain in a permanent habitable zone (PHZ) to allow liquid water to exist near its surface. For a planet in a PHZ, the values of the insolation function remain within the HZ. Based on the Wiliams & Pollard (2002) argument that a planet's atmosphere and oceans can buffer against changes in sunlight, as long as the average duration of insolation is within the HZ, the extreme value of insolation can be ignored. We call this averaged habitable zone (AHZ). Formally the combined spectrally weighted insolation on the planet is defined as:

$$\mathbb{S}_{I,O}(t) = \frac{L_A}{SA_{I,O}} a^{-2}(t) + \frac{L_B}{SB_{I,O}} b^{-2}(t) \qquad (2)$$

where *a* and *b* are the distance between the planet and star A and the planet and star B, in addition, subscripts *I, O* represent the inner and outer edge of the HZ.

$$PHZ: \max(\mathbb{S}_I) \leq 1 \qquad \text{and} \qquad \min(\mathbb{S}_O) \geq 1$$

$$AHZ: \ \langle \mathbb{S}_I \rangle \leq 1 \qquad \text{and} \qquad \langle \mathbb{S}_O \rangle \geq 1 \qquad (3)$$

where $\langle \mathbb{S} \rangle$ denotes the time-averaged combined stellar insolation. Note that all dynamically informed HZ neither depend on angular variables nor on time. Therefore, PHZ and AHZ form concentric rings around the center. The detailed procedure for the above equations is given in the paper (Eggl et al. 2020), which include the code to calculate the HZ, we can quickly calculate the values of the above simulation samples in section 2.1.

The evolution of life on a planet requires not only that the initial conditions of the planet are favorable to the evolution of life, but also that the planet subsequently remains habitable without interruption. Figure 1 indicates that the orbit of the planet is stabilized in a circular region for a long time because of the gravitational perturbations of the circumbinary on the planet. Therefore, these planetary systems can be divided into three classes: habitable, part-habitable and uninhabitable, based on planet trajectories and HZ. The purple ring is the planet's trajectory and the orange rings are the HZ inner and outer boundaries. If the planet's trajectory is within the HZ as the habitable system and if the planet's trajectory is outside the HZ as the uninhabitable system and if the planet's trajectory overlaps only partially with Hz as the part-habitable system. Meanwhile, this means that planet's trajectory is periodically in or out of the HZ, therefore, the life born in this environment may require it to go dormant periodically. Figure 2 shows the classification of planetary systems under the PHZ and AHZ conditions. Such classification criteria will continue to be applied to ML models.

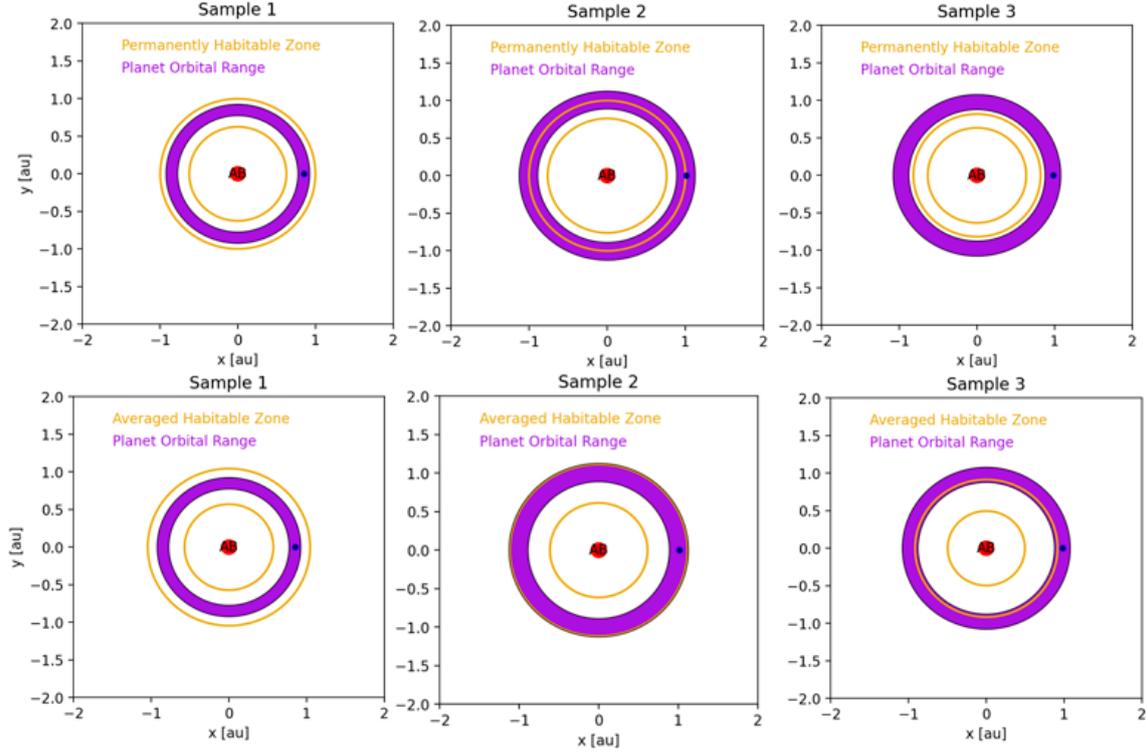

**Figure 2:** The purple ring is the planet's trajectory and the orange rings are the HZ inner and outer boundaries. The planetary system parameters are the same for the same column, but the HZ criteria are different, so the classification may be different.

## 3 Machine learning models

### 3.1 Deep neural networks

Deep neural networks (DNNs) have been demonstrated to be powerful tools in predictive applications, exoplanetary studies are no exception (e.g., Kipping & Lam 2017; Lam & Kipping 2018). In our work, we used Pytorch which used a syntax similar to Python NumPy to describe models.

### 3.2 Prepare the training data

In the section 2.1, we obtained three datasets representing the different mass of the planet ($m_p = 0, m_\oplus, m_{jupiter}$) by N-body simulation. Each dataset can be divided into two categories according to the stability of planet's orbit: "stable" and "instable". Figure 3 shows the distribution of the earth mass dataset, $m_p = m_\oplus$.

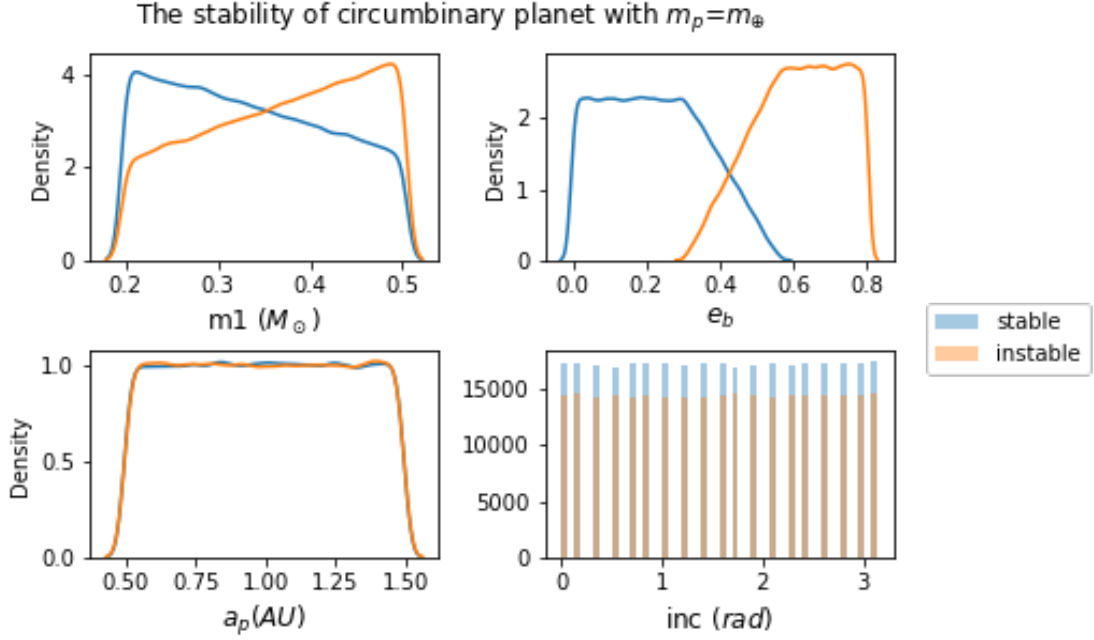

**Figure 3:** There are 6*10⁵ data in the figure with $m_p = m_\oplus$. There are three kernel density estimate (KDE) subgraphs of secondary star's mass, $m_2$, eccentricity of binary stars, $e_b$, planet's semimajor axis, $a_p$, and one histogram of the inclination of the planet, $i$. There are two colors to distinguish the orbital status, as indicated on the legend.

Next, we focus on the HZ of these stable samples. To calculate the "dynamically informed habitable zones", we need to add new parameters in these stable samples, such as the luminosity and surface temperature of binary stars, $L_A, L_B, T_A, T_B$. To get results similar to those of observed samples and not rely on the stellar model, we use Kelper CBP samples to find the linear relationship between stellar mass, surface temperature and stellar radius. These Kelper CBPs are listed in Table 2.

The following equation gives a calculation of the stellar luminosity:

$$\frac{L}{L_\odot} = \left(\frac{R}{R_\odot}\right)^2 \left(\frac{T_{eff}}{T_\odot}\right)^4 \qquad (4)$$

where $T_\odot = 5800K$

Besides, Figure 3 illustrate that the inclination of the planet has little effect on the statistical distribution of these samples, so we ignore this parameter in later studies. Eventually, each sample in the dataset will contain the following parameters: $m_A, m_B, T_A, T_B, L_A, L_B, e_b, a_p$. Combining the PHZ and AHZ calculations in the section 2.2 and classification criteria for planetary systems, we obtain the training dataset for ML algorithm.

**Table 2:** Physical parameters and orbital elements for the Kepler-16(AB), Kepler-

34(AB), Kepler-35(AB), Kepler-38(AB), Kepler-64(AB), Kepler-413(AB), Kepler-453(AB), Kepler-1647(AB) and Kepler-1661(AB) stellar binaries.

| System | $m_A(m_\odot)$ | $m_B(m_\odot)$ | $R_A(R_\odot)$ | $R_B(R_\odot)$ | $T_A(K)$ | $T_B(K)$ |
|---|---|---|---|---|---|---|
| Kepler-16 | 0.6897 | 0.20255 | 0.6489 | 0.22623 | 4450.0 | 3311.0 |
| Kepler-34 | 1.0479 | 1.0208 | 1.1618 | 1.0927 | 5913.0 | 5867.0 |
| Kepler-35 | 0.8876 | 0.8094 | 1.0284 | 0.7861 | 5606.0 | 5202.0 |
| Kepler-38 | 0.949 | 0.249 | 1.757 | 0.2724 | 5640.0 | 3325.0 |
| Kepler-64 | 1.528 | 0.378 | 1.734 | 0.408 | 6407.0 | 3561.0 |
| Kepler-413 | 0.820 | 0.5423 | 0.7761 | 0.484 | 4700.0 | 3463.0 |
| Kepler-453 | 0.944 | 0.1951 | 0.833 | 0.2150 | 5527.0 | 3226.0 |
| Kepler-1647 | 1.210 | 0.975 | 1.7903 | 0.9663 | 6210.0 | 5770.0 |
| Kepler-1661 | 0.841 | 0.262 | 0.762 | 0.276 | 5100.0 | 3585.0 |

### 3.3 Network architecture

In our work, we used the DNNs to predict the habitable zones of circumbinary planets. In the section 3.2, we have prepared the training dataset. Then, we take a set of N-dimensional ($m_A, m_B, T_A, T_B, L_A, L_B, e_b, a_p$) inputs and pass the data through 'hidden' layers of neurons, in which on-linear activation functions transform the data. The output of the network is compared with the true label (habitable, part-habitable and uninhabitable), and the error calculated via loss function. Next, we used the back-propagation algorithm to adjust the weights. Finally, we find the minimum value of the loss function through iterative operations.

Neural network architecture can be determined using the following parameters: the number of hidden layers, the number of neurons in these hidden layers, the activation functions, the loss function, and dropout. In our work, the net's architecture is set up by 5 hidden layers of 64 neurons each, using ReLU (rectifier linear unit) activation function, CrossEntropyLoss loss function, and SGD optimizer. The ReLU activation function, which lives in hidden layers, is defined as

$$f(x) \equiv \max(0, x) \quad (5)$$

The SGD optimizer reduces computational cost at each iteration and is a good estimate of the gradient descent. Dropout is a technique used to avoid overfitting, an issue endemic to models as complex as DNNs (Srivastava et al. 2014).

### 3.4 Feature selection and Iterative learning

In section 3.2, we get three training datasets for each mass of the planet ($m_p = 0, m_\oplus, m_{jupiter}$), and each group has eight parameters and two labels (PHZ and AHZ). Then we select $m_A, m_B, T_A, T_B, L_A, L_B, e_b, a_p$ as features to train our DNNs. Before starting, we need to pre-process the data. For these continuous values, $m_A, m_B, T_A, T_B, L_A, L_B, e_b, a_p$ , we need put them on a common scale. The standardization is defined as

$$x = (x - x_{min})/(x_{max} - x_{min}) \tag{6}$$

Therefore, these features are normalized between 0 and 1. The two labels are types of planetary orbits according to the PHZ and AHZ standards, respectively. Therefore, we train the DNNs in different HZ models, separately.

When training a DDN, we hope to transfer the data set iteratively through the network, and constantly adjust the weight of the neural network to minimize the loss function. To do this, we put data sets into training and validation sets that we use 9:1 to adjust parameters, select features, and make other decisions regarding the learning algorithm. Next, we train the network in rounds of "epoch" until the loss on the set of validations ceases to be significantly reduced. An 'epoch' means that all training examples pass through the network and the subsequent back-propagation. We set the 'epoch' to 100, and there are some other parameters such as: batch size = 1024, learning rate = 0.03. In the next section, we will discuss the result of the REBOUND and DNNs.

## 4. Results.
### 4.1 The stability of non-coplanar circumbinary planetary systems

We analyzed the statistical distribution map of the Earth mass samples, $m_p = m_\oplus$. Figure 3 illustrates that the samples of stable are mainly concentrated in the area with the smaller second star's mass, $m_2$, and smaller eccentricity of binary, $e_b$; the samples of instable planets are mainly concentrated in the area with larger second star's mass, $m_2$, and larger eccentricity of binary, $e_b$. In addition, we find that the initial semi-major axis, $a_p$, and inclination $i$, of the planet have little effect on the stability distribution of the samples in our model. This conclusion is quite different from previous studies (Lam & Kipping 2018; Chen, C., et al 2019). The main reason for these differences is that we focus on planetary systems near the HZ, therefore the planets are further away from the binary stars in our model. Figure 4 and 5 show the sample statistical distribution of Jupiter's mass and zero mass, $m_p = m_{jupiter}, and\ m_p = 0$. This indicates that the mass of the planet is not a major factor affecting the stability of the system in our model. Therefore, the following studies mainly focus on Earth mass, Jupiter mass and zero mass samples as shown in the Appendix section.

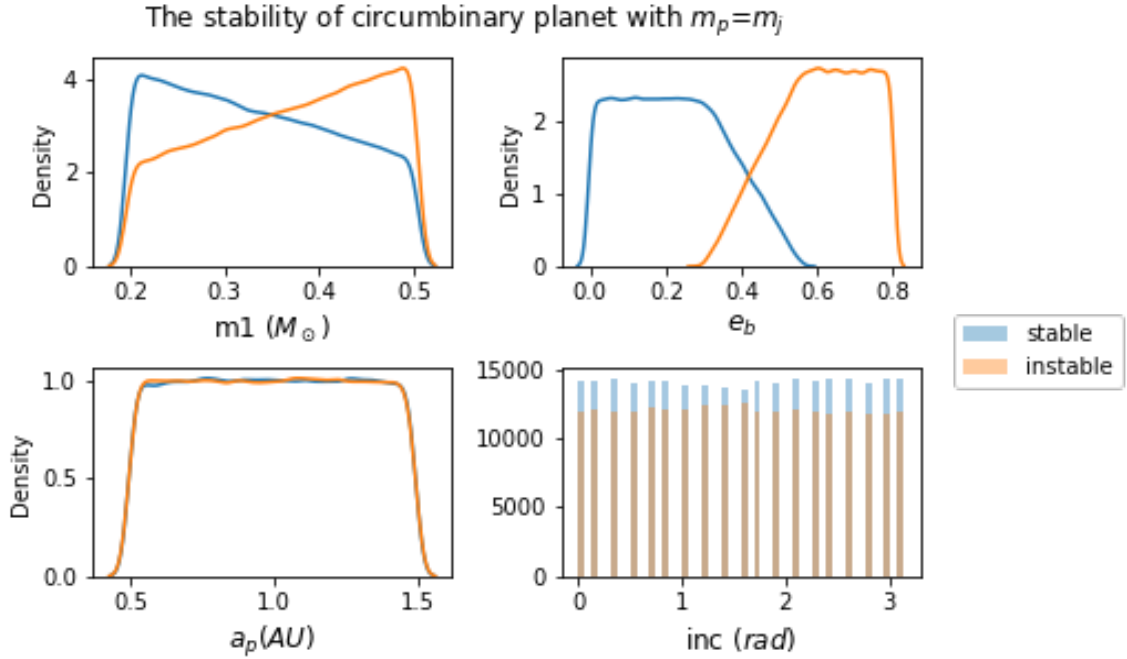

**Figure 4**: Kernel density estimate (KDE) of stable and unstable planetary orbits as a function of orbital and binary parameters. There are 496007 data in the figure with $m_p = m_{jupiter}$, secondary star's mass, $m_2$, eccentricity of binary stars, $e_b$, planet's semimajor axis, $a_p$, The fourth panel shows a stacked histogram of the stability as a function of the inclination, i, of the planet.

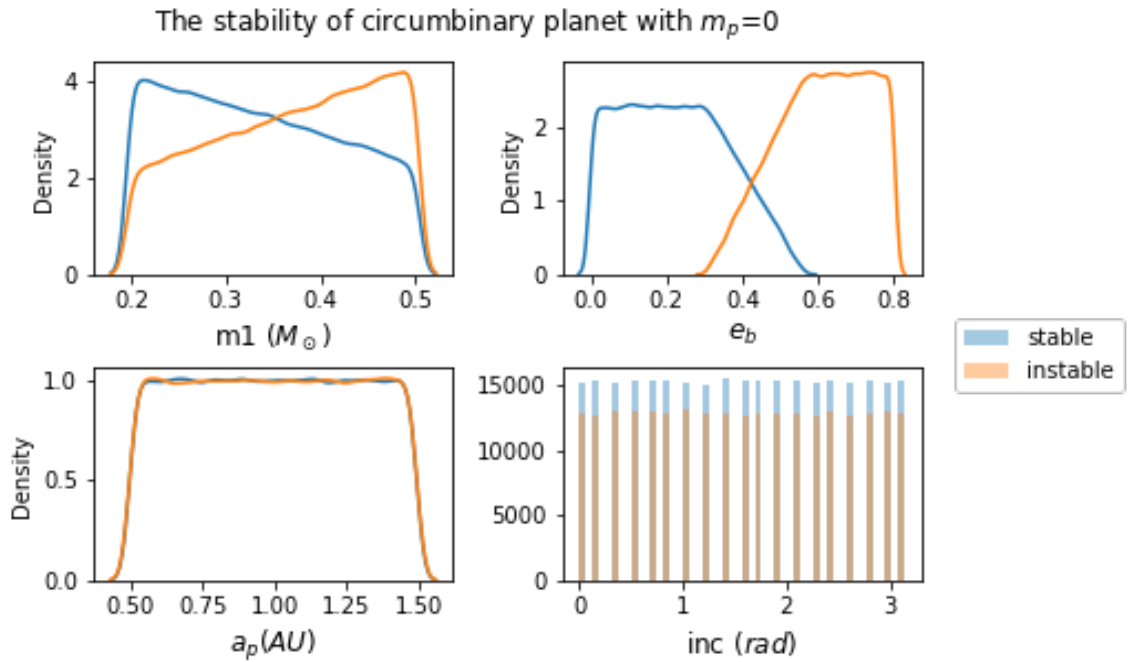

**Figure 5:** Kernel density estimate (KDE) of stable and unstable planetary orbits as a function of orbital and binary parameters. There are 534913 data in the figure with $m_p = 0$, secondary star's mass, $m_2$, eccentricity of binary stars, $e_b$, and planet's semimajor axis, $a_p$. The fourth panel shows a stacked histogram of the stability as a function of the inclination, *i*, of the planet.

## 4.2 Dynamically informed habitable zones of CBPs

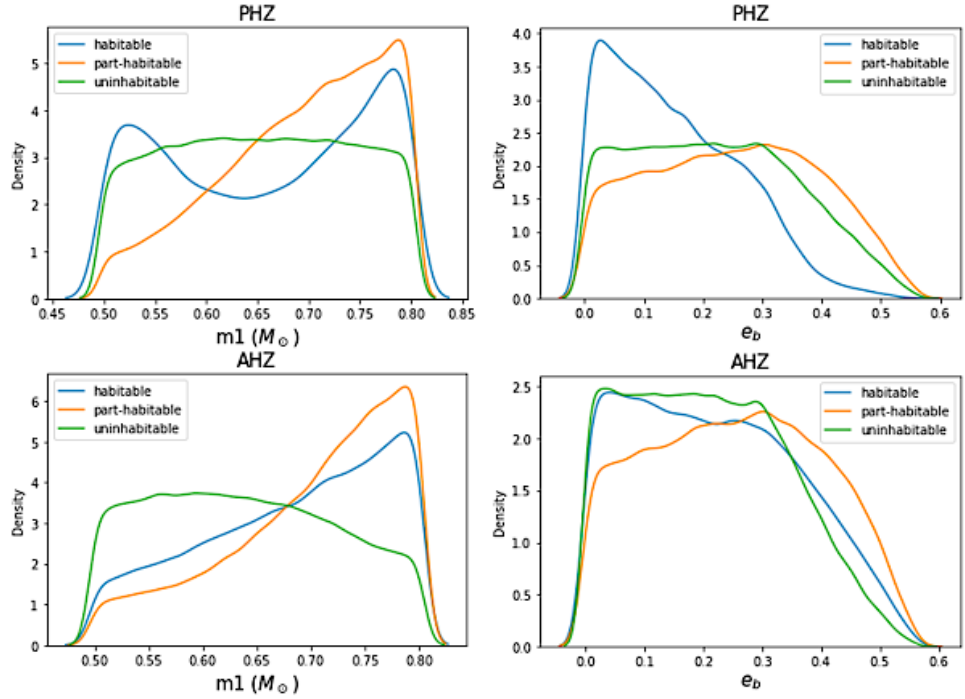

**Figure 6:** Kernel density estimate (KDE) of the planetary system samples under PHZ(top) and AHZ(bottom) conditions. The left-hand figure shows the distribution of the mass of the primary star and the different orbital types of the planet. The right-hand figure shows the distribution of the eccentricity of the binary star and the different orbital types of the planet.

We described the concept of "dynamically informed habitable zones of CBPs" and the calculation methods for PHZ and AHZ, in Section 2.2. These planetary systems can be divided into three classes: habitable, part-habitable and uninhabitable, based on planet trajectories and HZ's conditions (PHZ or AHZ). Figure 6 shows the distribution of different habitable types of planetary systems under PHZ conditions. In this case, we do not consider the influence of planet's climate inertia on stellar insolation, which illustrates that the planet's trajectory is more likely to be in the HZ completely, when the mass of the binary is similar or the mass of the primary star is dominant. Moreover, when the eccentricity of the binary star is low, the planet's trajectory is more likely to be in the HZ totally. However, under the AHZ condition, the distribution of planetary system samples is significantly different. Moreover, considering the inertia of the planet's climate, the orbit of the planet can still be entirely located within the HZ when the binary star system has a certain eccentricity. Furthermore, the figure7 shows the dependence of habitability on the initial semi-major axis. In this case, the distributions of the uninhabitable samples under PHZ and AHZ conditions are consistent, with the peaks appearing too close or too far away from the binary stars. However, under the PHZ condition, the peaks of the sample distributions of habitable and part-habitable are staggered, but the distribution peaks of samples habitable and part-habitable are the same under the AHZ condition.

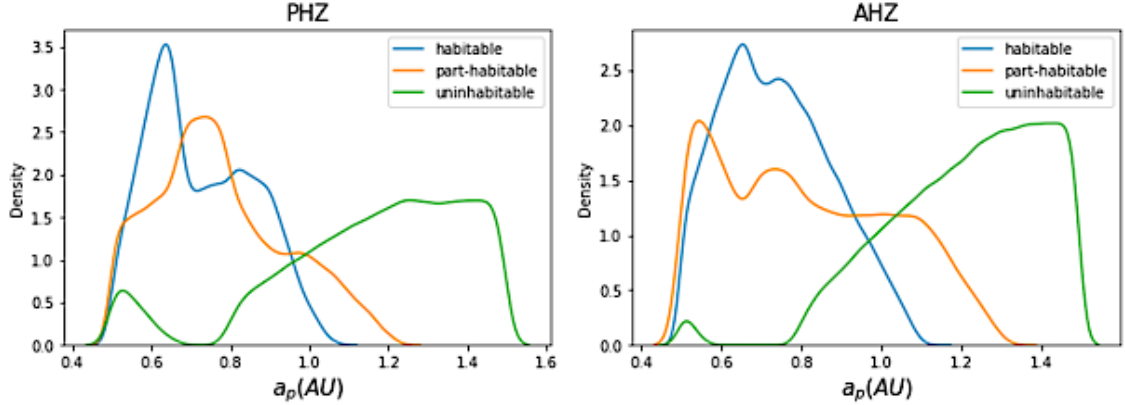

**Figure 7:** Kernel density estimate (KDE) of the planetary system samples under PHZ (left) and AHZ (right) conditions. These two figures show the distribution of the planet's initial semi-major axis and the different orbital types of the planet.

**4.3 The performance of machine learning**

In this paper, we consider the influence of the HZ on both binary stars and planetary orbits. As a result, our models of planetary systems become more detailed and complex. Therefore, when the amount of sample data is large, we need to use a machine learning algorithm to improve the classification speed. Normally, the performance of ML is related to the size of trainset data, the more data the higher accuracy but using N-body simulations to generate the trainset data is computationally expensive. Hence, we choose the samples about $3*10^5$ to train the Deep Neural Networks (DNNs). Once training is complete, we can use DNNs to quickly predict more samples, that is an important advantage of ML algorithms.

Next, we split the dataset 1/10 into test and training sets. The trainset data is used to train the structure of the neural network, so we save the model parameters of DNNs, and then using the test- set test the performance of DNNs. In this work, we use accuracy to evaluate the performance of machine learning models. Here is the definition:

$$acc = T/T + F$$

where T is the number of samples that are correctly predicted, and F is the number of samples where the prediction was wrong. Table 3 shows the results of DNNs on PHZ and AHZ.

**Table 3:** The performance of machine learning algorithms

| Dynamically informed habitable zones | Accuracy |
|---|---|
| Permanent habitable zone (PHZ) | 99.19% |
| Averaged habitable zone (AHZ) | 99.12% |

## 5. Summary and Discussion

With the development of transit timing variations (TTVs), more and more multi-planetary systems have been found. This has greatly expanded our understanding of the diversity of exoplanets, and has led to the scientific study of complex planetary systems. Recently, many studies have focused on non-coplanar planetary systems. (Doolin & Blundel 2011; Chen et. al 2020) and dynamically informed habitable zones (Eggl et al. 2012; Eggl et al. 2020). In this paper, therefore, we combine the latest research advances in these two areas to study the HZ of CBPs based on planetary trajectory and dynamically informed habitable zones. This makes sense because the birth and evolution of life required planets to remain habitable for long periods of time. In particular, the evolution of intelligent life on a planet requires not only that the initial conditions of the planet are favorable to the evolution of life, but also that the planet subsequently remains habitable without interference. In this paper, we consider both planetary trajectories and planetary climate inertia, and divide the habitability of planetary systems into three categories: habitable, part-habitable and uninhabitable. With the development of telescope technology, we will discover more binary star systems, and our work will be helpful to find the planet systems suitable for life from these binary star systems. In addition, the more complex our model is, the more computational resources are needed. Therefore, we recommend the DNN method to quickly and efficiently classify different planetary systems.

In conclusion, our work shows that using this model, the mass and inclination of the planets are not the main factors affecting the stability of their orbits, but the mass ratio of the binary stars and the eccentricity of their orbits. In addition, these stable CBP samples, the trajectory of the planet is constantly changing to form a ring, on long time scales, therefore, the HZ orbital types can be divided into three categories, and dynamically informed habitable zones are further divided into PHZ and AHZ. In this case, the mass ratio and orbital eccentricity of binary stars are still the main factors, but the habitability distribution of planets are very different under different HZ standards (PHZ or AHZ), see Figure 6. There are aspects of our model that can be improved in the future, such as increasing the simulation period of planetary systems, improving the climate inertia model of planets and extending it to more types of binary systems such as circumstellar planetary systems. Our code is viewable on GitHub and can be accessed through the following link: https://github.com/kong996/ML_CBP_HZ.git .

## 6. Acknowledgement

Authors JHJ and KAF were supported by the Jet Propulsion Laboratory, California Institute of Technology, under contract with NASA. We also acknowledge NASA ROSES XPR program for support. Author RB acknowledges the financial support from the SNSF under grant P2BEP2_195285. Authors ZK and ZZ thanks the support by the National Natural Science Foundation of China under Research Grants No. 11633001, 11920101003 and 12021003, and the Strategic Priority Research Program of the Chinese Academy of Sciences, Grant No. XDB23000000. We thank Xiaoming Jiang for constructive comments and discussions during development of this study and thank Bingkun Li for helping to optimize the ML algorithms.

**Data availability:** The exoplanet data used for this study can be downloaded at the NASA Exoplanet Archive (https://exoplanetarchive.ipac.caltech.edu/). The simulation codes can be downloaded at https://github.com/kong996/ML_CBP_HZ.git. For additional questions regarding the data sharing, please contact the corresponding author at Jonathan.H.Jiang@jpl.nasa.gov.

**References**

Armstrong, D.J., H. P. Osborn, D. J. A. Brown et al., 2014, MNRAS, 444, 1873, doi: 10.1093/mnras/stu1570.
Chen, C., A. Franchini, S. H. Lubow et al., 2019, MNRAS, 490, 5634, doi: 10.1093/mnras/stz2948.
Cuntz, M., 2014, ApJ, 780, 14, doi:10.1088/0004-637X/780/1/14.
Doolin, S., & K. M. Blundell 2011, MNRAS, Volume 418, Issue 4, 2656–2668, doi: https://doi.org/10.1111/j.1365-2966.2011.19657.x
Doyle, L. R., J.A. Carter, D.C. Fabrycky, et al. 2011, Science, 333, 1602, doi: 10.1126/science.1210923.
Eggl, S., E. Pilat-Lohinger, N. Georgakarakos et al., 2012, ApJ, 752, 74, doi:

10.1088/0004-637X/752/1/74.

Eggl, S., N. Georgakarakos, E. Pilat-Lohinger, 2020, Galaxies 8, 65. doi:10.3390/galaxies8030065.

Haghighipour, N., & L. Kaltenegger, 2013, ApJ, 777, 166, doi: http://dx.doi.org/10.1088/0004-637X/777/2/166.

Holman, M. J., & P.A. Wiegert, 1999, AJ, 117, 621, doi: 10.1086/300695

Kipping, D. M., & C. Lam, 2017, MNRAS, 465, 3495, doi: 10.1093/mnras/stw2974.

Kostov, V.B., J. A. Orosz, W. F. Welsh et al., 2016, ApJ, 827, 86, doi: 10.3847/0004-637X/827/1/86

Lam, C., & D. Kipping, 2018, MNRAS, 476, 5692, doi: 10.1093/mnras/sty022.

Moe, M., and R. D. Stefano, 2017, ApJ, Volume 230, Issue 2, article id. 15, 55 pp, doi: 10.3847/1538-4365/aa6fb6.

Orosz, J.A., W.F. Welsh, N. Haghighipour et al., 2019, AJ, 157, 174, doi: https://doi.org/10.3847/1538-3881/ab0ca0.

Rabl, G., and R. Dvorak,1988, A&A, 191, 385.

Raghavan, D., H. A. McAlister, T.J. Henry et al., 2010, ApJ, Volume 190, Issue 1, pp. 1-42, doi: 10.1088/0067-0049/190/1/1.

Rein, H., 2012, Astro-ph.EP, doi: https://arxiv.org/abs/1211.7121v2

Rein, H., S-F. Liu, 2012, A&A, 537, A128, doi: 10.1051/0004-6361/201118085.

Rein, H., & D. S. Spiegel, 2015, MNRAS, 446, 1424, doi: 10.1093/mnras/stu2164.

Socia, Q. J., W. F.Welsh, J. A. Orosz, et al. 2020, AJ, 159, 94, doi: https://iopscience.iop.org/article/10.3847/1538-3881/ab665b

Srivastava N., G. Hinton, A. Krizhevsky, I. Sutskever, R. Salakhutdinov, 2014, Journal of Machine Learning Research, 15, 1929.

Welsh, W.F., J. A. Orosz, D. R. Short et al., 2015, ApJ, 809, 26, doi: http://dx.doi.org/10.1088/0004-637X/809/1/26

Williams, D. M. & D. Pollard, 2002, Astrobiol ,1, 61–69, doi: https://doi.org/10.1017/S1473550402001064.

# Appendix

**The planetary system samples of Jupiter mass**

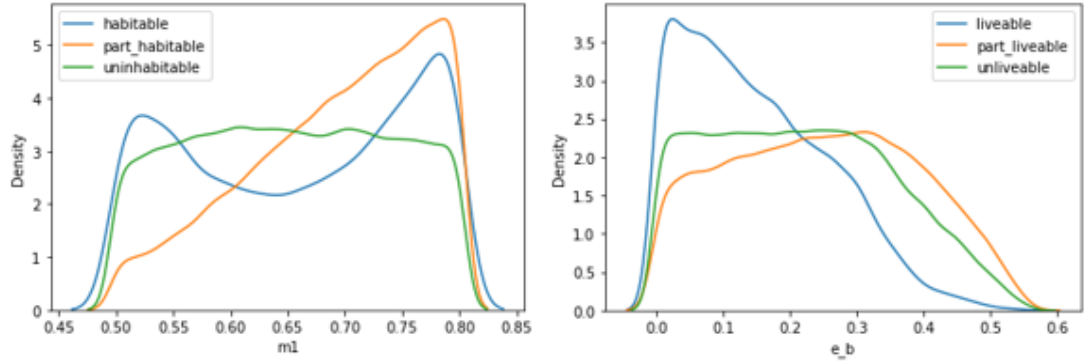

**Figure 8:** The figure shows the kernel density estimate (KDE) of the $m_p = m_{jupiter}$ planetary system samples under PHZ conditions. The left-hand figure shows the distribution of the mass of the primary star and the different orbital types of the planet. The right-hand figure shows the distribution of the eccentricity of the binary star and the different orbital types of the planet.

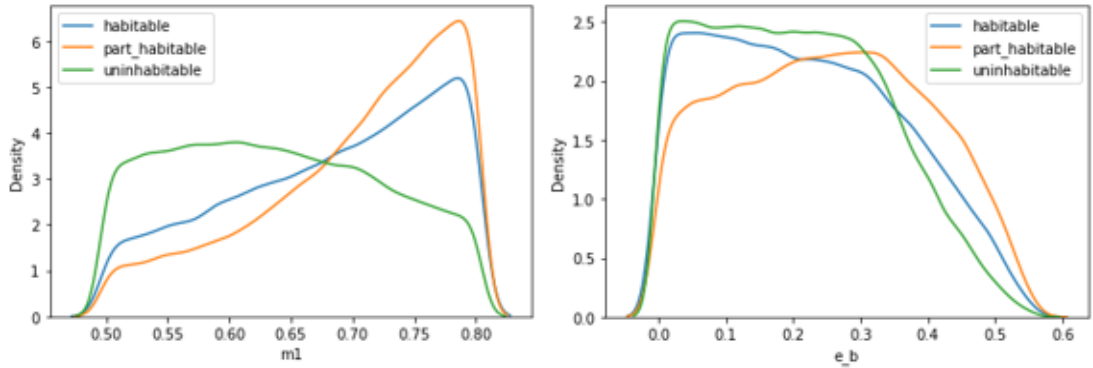

**Figure 9:** The figure shows the kernel density estimate (KDE) of the $m_p = m_{jupiter}$ planetary system samples under AHZ conditions. The left-hand figure shows the distribution of the mass of the primary star and the different orbital types of the planet. The right-hand figure shows the distribution of the eccentricity of the binary star and the different orbital types of the planet.

**The planetary system samples of zero mass**

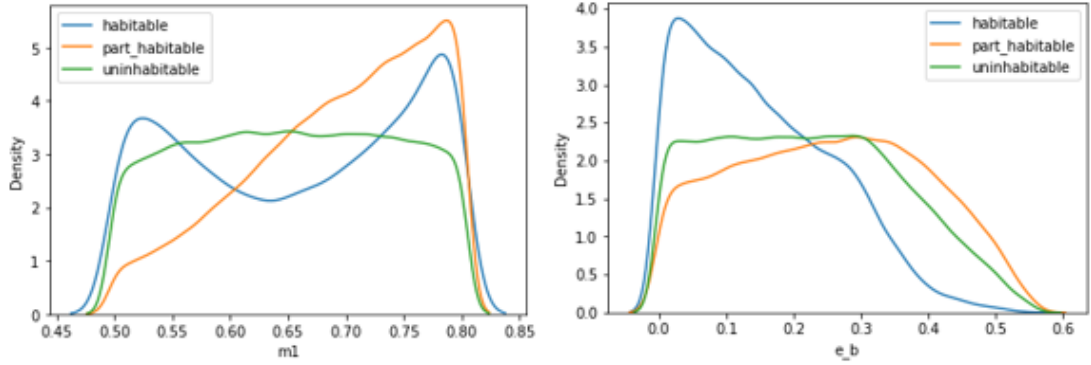

**Figure 10:** The figure shows the kernel density estimate (KDE) of the $m_p = 0$ planetary system samples under PHZ conditions. The left-hand figure shows the distribution of the mass of the primary star and the different orbital types of the planet. The right-hand figure shows the distribution of the eccentricity of the binary star and the different orbital types of the planet.

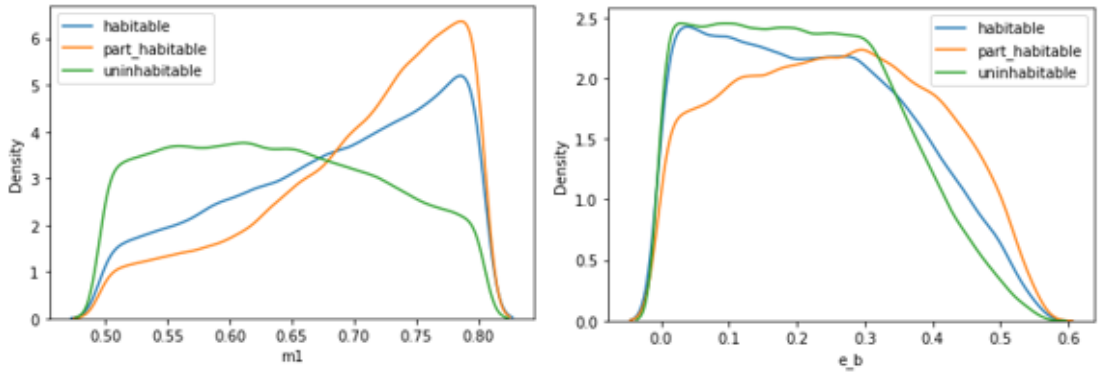

**Figure 11:** The figure shows the kernel density estimate (KDE) of the $m_p = 0$ planetary system samples under AHZ conditions. The left-hand figure shows the distribution of the mass of the primary star and the different orbital types of the planet. The right-hand figure shows the distribution of the eccentricity of the binary star and the different orbital types of the planet